\documentstyle[aps,preprint]{revtex}
\begin{document}
\draft
\title{ NMR  in ``underdoped'' and
``overdoped'' YBa$_2$Cu$_3$O$_{7-\delta}$ compounds:
Fermi-liquid approach. }
\author{ D.N. Aristov and A.G. Yashenkin}
\address{ Petersburg Nuclear Physics Institute, Gatchina,
St.Petersburg 188350, Russia }
\date{\today}
\maketitle
\begin{abstract}
We examine the NMR experimental data for the normal state
of YBa$_2$Cu$_3$O$_{7-\delta}$  within the Fermi-liquid
approach. We show that the observed temperature dependence
of the Knight shift and $^{17}$O ($^{89}$Y) relaxation rate
in ``underdoped'' ( $T_c=60\,$K ) compound can be
interpreted as resulting from a peak in the density of
states near the Fermi level.  The possible origin of such
peak is the quasi-one-dimensional van Hove singularity in
the fermionic spectrum, conjectured in \cite{abricamp}.
The proposed spectrum allows us to account for the
qualitatively different NMR data for ``underdoped''
and ``overdoped''( $T_c=90\,$K ) compounds by varying the
value of chemical potential only.
\end{abstract}

Last years for the description of properties of
high-$T_c$ cuprates a series of theoretical
concepts have been proposed, invoked to explain
the high-$T_c$ phenomenon itself and the anomalies of the
normal state of these compounds. We wish to mention here the
van-Hove singularity (vHs) scenario \cite{Mark} and nearly
antiferromagnetic Fermi liquid (NAFL) \cite{NAFL} model.

In this paper we address mainly to the NMR experiments in the
normal state of underdoped compound YBa$_2$Cu$_3$O$_{6.6}$
\cite{taki-CuOY,horva}  ( $T_c\approx 60\,$~K ), which
exhibits the properties fairly distinctive from those of
overdoped YBa$_2$Cu$_3$O$_{6.95}$.  The main difference
relates to the temperature dependence of the Knight shift
above $T_c$. It has been experimentally observed
\cite{taki-CuOY,horva}, that the Knight shift $K$ in
underdoped compound increased with temperature and
apparently saturated at $T\sim 300\,$~K , while it should
be constant for ordinary Fermi liquid. The $^{17}$O and
$^{89}$Y relaxation rates $(TT_1)^{-1}$ behaved similarly
to the Knight shift \cite{NAFL,taki-CuOY}, while the copper
relaxation rate $^{63}(TT_1)^{-1}$ suffered a maximum at
$T\approx 140\,$~K.

It was shown \cite{NAFL} that the qualitative difference in
the relaxation rates at different sites, as well as the
unusual proportionality of $(TT_1)^{-1}$ to
$K$ in the underdoped compound can be understood in the
frame of semiphenomenological NAFL approach.  However, the
Knight shift was treated as a parameter of the theory, i.e.
has not been explained.  Another parameter, the
antiferromagnetic correlation length essentially varied
with temperature, which clearly contradicts to the
inelastic neutron scattering data (INS) \cite{INS}. Next,
this theory can not be naturally continuated into the
superconducting temperature region.

The temperature dependence of the Knight shift as well as
some of INS data for the underdoped materials are usually
referred as the opening of spin-gap with lowering of the
temperature \cite{horva}.

In the present paper we explain the
observed temperature dependence of the Knight shift and
O ( Y ) relaxation rate for YBa$_2$Cu$_3$O$_{6.6}$ compound
within the Fermi liquid approach.  Our
consideration utilizes the hypothesis of the
peak in the density of states (DOS) near the Fermi level in
this system, which idea is shared by the vHs models
\cite{Mark}.  This idea has been previously discussed in
connection with the thermodynamic measurements in YBaCuO
compound \cite{Tsuei-PRL}.  On the other hand, the recent
photoemission data witness in favor of flat parts of
electron spectrum near the Fermi surface
\cite{abricamp,Y-Ba-ARPES} in high-$T_c$
cuprates, which should correspond to the peak in DOS at
energies close to the Fermi level.

Leaving apart the widely discussed question whether
it is possible to explain high $T_c$ in the frame of the
peak-in-DOS hypothesis, we confine ourself with the
consequences for the normal state quantities observable in
NMR experiments.

Our main idea is as follows. The values of the spin
part of Knight shift $K$ and the reduced nuclear
spin-lattice relaxation rate $(TT_1)^{-1}$ (RR) in
metals are determined by the electrons, whose energies
( measured from the Fermi level ) are of order of temperature.
If the density of states within this interval is nearly
constant one obtains the temperature independent values of
$K$ and $(TT_1)^{-1}$.  However, if the DOS function
possesses an additional (small) energy scale near the Fermi
level, it might result in the essential variation of these
quantities with temperature. We found that the experimental
data for the Knight shift for YBa$_2$Cu$_3$O$_{6.6}$ are
consistent with a peak in DOS at the energy $\sim 40\,
$meV.  On the other hand, the relaxation rate is determined
by the integrated square of DOS, in which quantity the peak
is reproduced as well. So one can conclude that the
temperature dependence of $(TT_1)^{-1}$ should be similar
to one of $K$ in accordance with experimental data for
$^{17}$O and $^{89}$Y. Discussing the origin of such peak
in DOS, we argue that the most likely reason for it could
be the quasi-one-dimensional vHs conjectured previously by
Abrikosov, Campuzano and Gofron \cite{abricamp} on the base
of photoemission data.

Moreover, we show that the proposed DOS function
permits us to account for the qualitative difference
between the compounds with $T_c\approx60\,$K and
$T_c\approx90\,$K by the difference in the chemical
potentials only.


{\em Knight shift}.
The spin part of the Knight shift is proportional to the
static susceptibility of electron system. In the normal state it
can be represented through the DOS function $\rho(x)$ in
the form \cite{slichter}:

\begin{equation}
 K^\alpha = \frac12 A^\alpha(0) \frac{\gamma_e}{\gamma_n}
\int \frac{dx \, \rho(x)}{4T\cosh^2((x-\mu)/2T)}
\label{knight1}
\end{equation}

\begin{equation}
 \rho(x)  = \sum_k \delta(E_k -x) |u_k|^2
\label{rho}
\end{equation}
here $A^\alpha({\bf q})$ is the hyperfine coupling
constant, $\gamma_e$ ($\gamma_n$) is the electron (nucleus)
gyromagnetic ratio and $\mu$  is the chemical potential.
The quasiparticle weight $|u_k|^2$ in (\ref{rho}) includes either
the amplitude of Bloch function on the nucleus ( for
$s$-electrons ) or the coefficient in the diagonalizing
Hamiltonian ( for multi-band electronic structure ) or the
effects of carriers' interaction.  Our main assumption
henceforth is small dependence of $u_k$ ( and therefore
$\rho(x)$ ) on $T$ \cite{fnote}.

The usual assumption is that the density of states is
almost constant near the Fermi surface. It is
valid, e.g., for the  Fermi gas, when $E_k=
k^2/(2m)$ and $u_k=1$ so that $\rho(x) = \rho(E_F)(1+
O(x/E_F))$. Hence, for the typical values of the
Fermi energy $E_F=\mu\sim 1\,$~ eV the temperature
corrections to $K^\alpha$ are of order of $(T/E_F)^2\ll1$.

If however $\rho(x)$ has some prominent feature at low
energy comparable with temperature, then the
temperature dependence of Knight shift would not be so
trivial. Namely, let us assume that $\rho(x)$ in
(\ref{knight1}) has a peak at $x=E_1 + \mu$. Then the
weight of this peak $\sim (T\cosh^2(E_1/2T))^{-1}$ rapidly
increases, as the temperature increases from zero to
$T\approx 0.6 E_1$, and starts to decrease slowly upon
further temperature raising.  Obviously the resulting
Knight shift would have the similar temperature behavior.

We present below a procedure which may let one judge
about spectral features from the Knight shift data.
We note that the expression (\ref{knight1}) can be
rewritten in terms of Mellin transforms in the following way :

\begin{equation}
K(s)= \rho_+(s)\varphi(1-s)   \label{meleqrho}
\end{equation}
Here we use the following definitions :

\begin{equation}
\begin{array}{rl}
\varphi(x) &= 2^{-1}\cosh^{-2}(x/2),\\
\rho_+(x) &=
\displaystyle
\frac{A(0) \gamma_e}{4\gamma_n}
(\rho(\mu+x)+\rho(\mu-x) )
\end{array}
\end{equation}
and
\begin{eqnarray}
f(s) &=& \int_0^\infty dx\,x^{s-1}f(x), \qquad
s=\sigma+i\lambda \\
f(x) &=& (2\pi i)^{-1}\int_{\sigma-i\infty}^{\sigma+i\infty}
ds\,x^{-s} f(s)
\end{eqnarray}

Of course, trying now to obtain from (\ref{meleqrho}) the
function $\rho_+$ one meets the difficulties common for the
badly-defined inverse problems, when $\varphi(1-s)\to0$.
Therefore we do not pretend to extract the exact form of
$\rho_+(x)$ and find instead only the regularized solution
$\rho_+(x,\delta)$ of the inverse problem in the form :

\begin{equation}
\rho_+(x,\delta) = (2\pi i)^{-1}\int_{\sigma-i\infty}^
{\sigma+i\infty} ds\,x^{-s} \frac{ K(s) \varphi^\ast(1-s)}{
|\varphi(1-s)|^2+\delta}
\end{equation}
where $\delta>0$ and $ \sigma $ is
chosen from the condition of simultaneous existence of
Mellin transforms $K(\sigma)$ and $\varphi(1-\sigma)$.  To
satisfy this condition, we subtract the constant term ( at
$T>300\,$ K ) from the observed temperature dependence of
Knight shift,  using the property $\int_0^\infty
dx\, \varphi(x) = 1$. We note that the restored function
$\rho_+(x,\delta)$ is almost insensitive to the choice of
$\sigma\in[0,1]$ and depends strongly on $\delta$. The
values of $\delta = (10^{-2} \div 10^{-4} )
|\varphi(1-\sigma)|^2$ give a shape of $\rho_+(x,\delta)$
with a visible details and , on the other hand, free from
the giant oscillations.

Choosing $\sigma=0.35$ and $\delta = 10^{-3}
|\varphi(1-\sigma)|^2$, we find the regularized solution
for symmetrical part $\rho_+$ of DOS which is shown in Fig.
\ref{Fig-DOS} by the dashed line. The resulting DOS has a
peak at $E_1\approx 40\,$~meV. Obviously the fine features
of DOS structure could not be restored during the procedure
of regularization so this picture delivers only the rough
information about the position and the general shape of the
DOS peak.


Let us discuss briefly the reasons of such strong variation
of DOS function at such a small energy scale.
First possibility relates
to the existence of vHs near the Fermi level.  This
possibility was widely discussed before, mainly for the
case of logarithmically divergent DOS,  $\rho(x) \sim
\ln(\Lambda/|x-E_1|)$.  This kind of vHs can be found
in a variety of papers based on the tight-binding
Hamiltonian and originates from the saddle point of
two-dimensional fermionic spectrum \cite{Mark,mfl}. We wish
to stress, however, that the logarithmic singularity is too weak
to produce the essential temperature variation of the Knight
shift (cf. \cite{mfl}).

At the same time, the recent photoemission
data for YBaCuO compounds \cite{abricamp,Y-Ba-ARPES} give a
strong confidence to the existence of the ``extended''
saddle point along the $\Gamma-Y$ symmetry direction.  It
has been shown \cite{abricamp} that the electronic spectrum
near the $(0,\pi)$ point of the 2D Brillouin zone can be
represented in the form \[E_k = t k_x^2 + E_1 ,\quad |k_y|
< k_0\approx .5 r.l.u.,\] where $ E_1 $  was found to be
within the experimental resolution of 30 meV.
The above
kind of spectrum results in DOS function of the form
\begin{equation}
\rho(x) = \rho_n + \rho_s(x),
\label{Eq:decom}
\end{equation}
 where $\rho_s(x)\sim
\rho_n \sqrt{tk_0^2/(x-E_1)}\, \theta(x-E_1)$, and $\rho_n\sim
t^{-1}$ is the smooth part of DOS.  The divergency in $\rho(x)$
is obviously smeared out either by the effects of higher
dimensionality or by the quasiparticle interaction. We model
these effects below by introducing the cutoff energy $E_c < E_1$.

We have found that the temperature dependence of the
observed Knight shift above $T_c$ \cite{taki-CuOY} can be well
fitted by (\ref{knight1}) with

\begin{equation}
\begin{array}{rl}
\rho(x+\mu) &\propto \left\{
\begin{array}{ll} 1&, x <  E_1\\
1+ \sqrt{\Lambda/ E_c }&, E_1 <  x <  E_1 +  E_c\\
\displaystyle 1 +
\sqrt{\frac{\Lambda}{x - E_1 }} &,
 E_1 + E_c < x ,
\end{array} \right.
\\
E_1 =& 25 {\rm meV}, \quad \Lambda = 1 {\rm eV}.
\end{array}
\label{modelDOS} \end{equation}
The cutoff energy $E_c$ was chosen to be 8 meV.
The corresponding function
$\rho_+(x)$ is also shown at Fig. \ref{Fig-DOS} by the
solid line.  The temperature dependence of the resulted
Knight shift is given at Fig. \ref{Fig-KW} ( the solid
curve (a) ).

Second possibility for the considered system to have
a small energy scale relates to the fact that the
O$_{6.6}$ compound is close enough to the metal-insulator
transition ( at the oxygen content O$_{6.4}$ ). It could
result in diminishing of the DOS value at the Fermi level
\cite{pseudogap}. In this case one can expect that the
small energy scale $E_g$ appears due to the strong
renormalization effects. However, if $E_g$ is formed in a
``dynamical'' way, one could expect that raising of the
temperature up to $E_g$ should lead to the considerable
changes in the quasiparticle weight and, consequently,
in the DOS shape. This point contradicts to our initial
assumption about essential temperature independence of
$\rho(x)$.

Until now we implied the $T-$ independent value of chemical
potential. It means the grand canonical ensemble description of
the considered system of CuO$_2$ planes. In other words, the
number of quasiparticles varies with temperature, which can be
realized by injecting the carriers from the Cu-O chains. It
appears to be reasonable in view of possible important role of
the chains in forming such one-dimensional kind of spectrum.
On the other hand, if one assumes the number of in-plane
carriers to be fixed, then the chemical potential is found
from the equation :
\begin{equation}
\int\frac{dx\,\rho(x)}{1+\exp((x-\mu)/T)}\, =\, const
\end{equation}
The consideration of the latter case with the above model
DOS shape (\ref{modelDOS}) leads to the temperature shift of
$\mu$ or, in other words, to the effective rescaling of
$E_1$. Hence, it can be shown that the considered
picture of the temperature variation of the Knight shift
persists in this case, too.


{\em The nuclear spin-lattice relaxation rate} $T_1^{-1}$
is determined by the imaginary part of the dynamical spin
susceptibility $\chi''({\bf q},\omega)$, integrated with
the formfactor over the Brillouin zone.  The INS experiments
\cite{INS} show the enhancement of $\chi''$ near the
AF wave-vector ${\bf Q}_0=(\pi,\pi)$.  These AF fluctuations are
apparently connected with the underlying physics, though the
detailed theory of their microscopic origin seems to be not
completed yet. It is however widely recognized, that the
different $T$-dependence of the relaxation rates for $^{63}$Cu
and $^{17}$O ($^{89}$Y) stems from the different formfactors,
filtering out the AF fluctuations at O (Y) sites.  Thus we
believe, that the copper RR is mostly determined by the
anomalous AF contribution which is beyond the scope of our
paper. On the other hand, we expect the oxygen (yttrium) RR
to be mainly determined by the ordinary Fermi-liquid
contribution.  We show below, that this latter expectation
is supported by the proposed picture of DOS, essentially
varying near the Fermi level.

To demonstrate it, we consider the NMR relaxation rate
$T_1^{-1}$ in the bare-loop approximation, i.e. ignoring the
effects of carriers' interaction.  Then in the case of magnetic
field along $\hat z$ one can write the following expression
\cite{slichter}:

\begin{equation}
(TT_1)^{-1} = \frac{\pi}{2\hbar} \int dx
\frac{\tilde \rho^2(x+\mu)} {4T\cosh^2(x/2T)}
\label{relrate2} \end{equation}

\begin{equation}
\begin{array}{rl}
\tilde \rho^2(x) = \sum\limits_{kq} &( |A^x({\bf q})|^2
+|A^y({\bf q})|^2) |u_k|^2 |u_{q+k}|^2
\\   &\times
\delta(x-E_k)
\delta(x-E_{q+k})
\end{array}
\label{rhosq}
\end{equation}

In the case of Fermi gas and isotropic contact hyperfine
interaction ( $A^\alpha({\bf k}) \equiv A$ ) one obtains $
\tilde \rho^2(x) = 2 A^2 \rho^2(E_F)$ and the well known
Korringa relation $(TT_1)^{-1} = 4\pi\hbar^{-1} (\gamma_n/\gamma_e)^2
K^2$. Now, when the $\rho(x)$ is decomposed onto the normal and
singular parts (\ref{Eq:decom}) , one can write

\[\tilde\rho^2(x) \sim \rho_n^2 + C_1 \rho_n\rho_s(x) +
C_2\rho_s^2(x), \] where the relative weights $C_1$ and $C_2$
are defined by the $q$-dependence of formfactors $|A^\alpha({\bf
q})|^2$ in (\ref{rhosq}).  It can be seen, that the term
$\rho_s^2(x)$ appears in (\ref{rhosq}), when both
wave-vectors {\bf k} and {\bf k+q} are near  the ``extended''
saddle point, so that the formfactors $A^{x,y}({\bf q})$ have
nearly zero argument. This observation allows us to drop the
$q$-dependence of formfactor in evaluating the relaxation rate,
at least at high temperatures, when $\rho_s^2$ plays a dominant
role.

After this remark we calculate RR as the function of
temperature, taking $\tilde\rho^2(x) = \rho^2(x)$. The
logarithmic divergence at $x= E_1  +\mu$ is cut off by the
energy $E_c$, discussed above. The obtained $T-$
dependence of RR is shown at Fig. \ref{Fig-KW} by the
dotted line (a).  One can see from this Figure that the
calculated relaxation rate is in qualitative agreement with the
experimental findings for $^{17}$O and $^{89}$Y
\cite{taki-CuOY,horva}.

At last, we note, that a strongly asymmetric shape of DOS
function (\ref{modelDOS}) provides a possibility to explain the
almost $T-$independent Knight shift in the `` overdoped ''
YBa$_2$Cu$_2$O$_{6.95}$ compound. Indeed, let us assume that the
main difference of ``90~ K'' compound from the ``60~ K'' one is
another value of the chemical potential or, equivalently,
another value of $E_1$ in (\ref{modelDOS}).

Then, taking  $E_1 < 0$, one can see from (\ref{knight1}),
(\ref{relrate2}) that i) at low temperatures the values of $K$
and $(TT_1)^{-1}$ are defined by the DOS values $\rho(\mu)
\propto 1+ \sqrt{\Lambda/|E_1|}$, and ii) at the higher
temperatures the region of relatively low DOS values comes into
play.

We calculate the temperature dependence of the Knight shift
and RR with (\ref{knight1}), (\ref{modelDOS}),
(\ref{relrate2}), similarly to the above ``underdoped'' case,
but now taking $E_1 = -25\,$meV. The results are shown at Fig.
\ref{Fig-KW} by the curves (b). One sees that in this case i)
raising of the temperature above $\sim |E_1|$ leads to
somewhat decrease of the Knight shift and RR, and ii) the
limiting values of the Knight shift and RR at high temperatures
should coincide for both over- and under-doped compounds, since
at $T\gg |E_1|$ the sign of $E_1$ is inessential.  Both these
statements are clearly supported experimentally
\cite{taki-CuOY,horva}.

Finally, addressing to the experiments with the underdoped
YBa$_2$Cu$_2$O$_{6.6}$ compound, we explain the temperature
variation of both Knight shift and oxygen (yttrium) NMR
relaxation rate within the frames of Fermi-liquid picture.
We argue that the observable $T-$ dependence of these
values can be understood if the $T-$independent density of
states in underdoped compound strongly varies with the
energy near the Fermi level.  ( At the same time the copper
relaxation, apparently determined by the AF fluctuations,
remains unexplained in our simple model. ) We show further,
that the qualitative difference between underdoped and
overdoped materials is accounted for the change of chemical
potential only, if the shape of DOS function is delivered
by the quasi-1D character of fermionic spectrum in YBaCuO
compounds, observed in the recent photoemission experiments
\cite{abricamp,Y-Ba-ARPES}.

\acknowledgements
We wish to thank Prof. S.V. Maleyev for valuable discussions
and critical reading of manuscript, M. Horvati\'c for
stimulating conversation, Yu. Lyanda-Geller for clarifying
comment. The research described in this publication was
made possible in part by financial support from the
International Science Foundation.

\begin{figure}
\caption{The function of DOS $\rho_+(x, \delta)$ restored
with regularization discussed in text (dashed line).
The model DOS function,
given by (\protect\ref{modelDOS}), is shown by the solid
line.} \label{Fig-DOS} \end{figure}

\begin{figure}
\caption{The calculated temperature dependences of the
Knight shift and $(TT_1)^{-1}$ with the model DOS function
given by (\protect\ref{modelDOS}).
(a) The ``underdoped'' case with $E_1 = 25\,$meV
 and  (b) the ``overdoped'' case with $E_1 = - 25\,$meV
} \label{Fig-KW} \end{figure}

\end{document}